\documentclass[aps,pra, twocolumn,a4paper,showpacs]{revtex4}
\usepackage{amssymb,amsmath,amsfonts,graphicx}
\usepackage{bm}

\newcommand{\ddps}{\frac{\partial}{\partial s}}
\newcommand{\ddpt}{\frac{\partial}{\partial t}}
\newcommand{\ud}{\mathrm{d}}
\newcommand{\Hs}{\mathbb{H}}
\newcommand{\Ls}{\mathbb{L}}
\newcommand{\Ks}{\mathbb{K}}
\newcommand{\Rs}{\mathbb{R}}
\newcommand{\Cs}{\mathbb{C}^2}
\newcommand{\un}{\openone}

\def\BE {\begin{equation}}
\def\EE {\end{equation}}
\def\BEA {\begin{eqnarray}}
\def\EEA {\end{eqnarray}}
\def\BES {\begin{subequations}}
\def\EES {\end{subequations}}
\def\BA {\begin{array}}
\def\EA {\end{array}}
\def\NN {\nonumber}
\def\ep {\epsilon}
\def\epnm {\epsilon^{2^{n-1}}}
\def\epn {\epsilon^{2^n}}

\def\epp {\epsilon^{2^p}}
\def\eppm {\epsilon^{2^{p-1}}}
\def\ad{{\rm ad}}

\def\rn{^{{\rm e}}_{n}}
\def\rnm{^{{\rm e}}_{n-1}}
\def\rnmm{^{{\rm e}}_{n-2}}
\def\rz{_{0}}
\def\ru{^{{\rm e}}_{1}}
\def\rd{^{{\rm e}}_{2}}
\def\rk{^{{\rm e}}_{k}}

\def\rp{^{{\rm e}}_{p}}
\def\rpm{^{{\rm e}}_{p-1}}
\def\ti{t_{{\rm i}}}
\def\tf{t_{{\rm f}}}

{
\begin{document}
\title{Unitary time-dependent superconvergent technique for pulse-driven quantum dynamics}
\date{\today}
\author{D.~Daems}\email{ddaems@ulb.ac.be}
\affiliation{Center for Nonlinear Phenomena and Complex Systems, Universit\'e Libre de Bruxelles,
CP 231, 1050 Brussels, Belgium}
\author{A.~Keller}\email{arne.keller@ppm.u-psud.fr}
\affiliation{Laboratoire de Photophysique Mol\'eculaire du CNRS,
Universit\'e Paris-Sud, B\^at. 210 - Campus d'Orsay,
91405 Orsay Cedex, France}
\author{S.~Gu\'erin}\email{sguerin@u-bourgogne.fr}
\author{H. R.~Jauslin}
\affiliation{Laboratoire de Physique de l'Universit\'e de Bourgogne,
UMR CNRS 5027, BP 47870, 21078 Dijon, France}
\author{O.~Atabek}
\affiliation{Same address as A. Keller}
\begin{abstract}
We present a superconvergent Kolmogorov-Arnold-Moser  type of perturbation theory for time-dependent Hamiltonians.
It is strictly unitary upon truncation at an arbitrary order and not restricted to periodic or quasiperiodic Hamiltonians.
Moreover, for pulse-driven systems we construct explicitly the KAM transformations involved in the iterative procedure.
The technique is illustrated on a two-level model perturbed by a pulsed interaction for which we obtain convergence all the way from the sudden regime to the opposite adiabatic regime.
\end{abstract}
%\keywords{ }
\maketitle
\section{Introduction}
The control of atomic and molecular dynamics by lasers has attracted considerable interest in the past decade.
Time-dependent systems are traditionally studied from a perturbative point of view with the Dyson expansion.
The two limiting cases of a sudden and an adiabatic switching of the perturbation have been extensively studied \cite{Galindo}.
In particular, this has resulted in the adiabatic theorem  and the superadiabatic expansion
\cite{Berry,Holthaus}.

We are interested in the case where the perturbation is switched on and off  on a time scale which need not be arbitrarily small or large.
It is well known that generally, upon truncation, the Dyson series for the evolution operator of a non-autonomous system is not unitary, giving rise to secular terms whose size grows with time.
For periodic and quasi-periodic perturbations, a number of schemes have been proposed \cite{Barata}, notably one
based on
 the Kolmogorov-Arnold-Moser (KAM) perturbation theory of classical
mechanics \cite{Gallavotti}.
Here we shall not be dealing with periodic or quasiperiodic systems but with time-dependent perturbations that are localized in time 
for which we develop, building on the KAM technique, a  unitary {\em superconvergent} 
perturbation theory.

The KAM iterative method  has been  introduced in quantum mechanics by Belissard      \cite{Bellissard} for periodic Hamiltonians.
It consists in generating at each step  (with the help of a unitary transformation) a new reference or {\em effective} Hamiltonian which
collects higher order terms of the perturbation that commute with the reference Hamiltonian
constructed at the preceding step.  
At the first iteration the order of the perturbation is reduced from $\ep$ to $\ep^2$ and, by considering the resulting Hamiltonian as a new starting point, the second transformation then reduces the order of the perturbation from $\ep^2$ to $\ep^4$.
Hence, at the $n$-th iteration the size of the remaining  perturbation is reduced from $\epnm$ to $\epn$.
The quantum  KAM technique has also been investigated for  periodic Hamiltonians by Combescure \cite{Combescure1,Combescure2} and more recently by Duclos and ~\v S\v tov\' \i cek \cite{Duclos}.
Quasiperiodic Hamiltonians have been considered by Blekher {\em et al.} in \cite{Blekher}.

All these authors have  implemented the KAM algorithm  in an {\em extended Hilbert space} constructed as the tensor product of the  Hilbert space in which the original Hamiltonian is defined and the space of square integrable functions on the circle.
This notion, introduced by Sambe \cite{Sambe} in the periodic case  and by Howland \cite{Howland} for more general time-dependent Hamiltonians, allows to construct a time-independent {\em extended Hamiltonian} (also called Floquet Hamiltonian in the periodic case) which is the starting point in the KAM algorithm.

The KAM iterative procedure requires solving two commutator equations  at each step. 
In \cite{SchererPRL,SchererII}  Scherer has shown, adapting ideas from classical mechanics going back to Poincar\'e, that these equations could be solved in terms of {\em time averages} of some operators related to the perturbation.

The generalisation of the KAM technique
to time-dependent Hamiltonians has been worked out by Scherer
\cite{SchererPLA,SchererIII}.
It has been built in close analogy to classical mechanics and involves an extended phase space which includes time as a coordinate and the energy of external sources as its conjugate momentum,
a notion  closely related to that of \cite{Howland}.
On the other hand, the KAM algorithm proposed by Scherer is quite cumbersome to use
and, in addition, is not guaranteed to yield a unitary evolution operator upon truncation.

In this paper we present a KAM algorithm for non-autonomous Hamiltonians   that is strictly unitary upon truncation at an arbitrary order.
Moreover, for pulse-driven systems we construct explicitly the KAM transformations and study the convergence of the algorithm on a specific case.

We start in Sec.~\ref{sub:KAM} by recalling the KAM technique and the quantum averaging method for time-independent Hamiltonians.
The notion of extended Hilbert space is presented in Sec.~\ref{sub:Kspace} at a purely formal level.
In Sec.~\ref{sub:time} we construct  a unitary KAM algorithm for time-dependent systems in an extended Hilbert space. 
In Sec.~\ref{sub:original} the quantum averaging technique is extended  to non-autonomous Hamiltonians in order to construct the KAM transformations directly in the original Hilbert space.
Sec.~\ref{sub:general} is devoted to perturbations that are switched on at some finite time in the past, for which
we calculate explicitly the successive time averages involved in the KAM algorithm.
We then focus on the case of pulse-driven two-levels systems, for which we 
resum exactly in Sec.~\ref{sub:resum} the infinite series of commutators yielding the remaining perturbations at a given step of the iterative procedure.
Finally, in Sec.~\ref{sub:results} the method is applied to a two-level system interacting with a sine-squared pulse,
taking the ratio of the characteristic duration of the pulse and the characteristic time of the free evolution as the small parameter $\ep$.
We show the remarkable result that  the KAM algorithm converges for all values of
the parameter $\epsilon$, even larger than unity, allowing to go from the sudden regime to the opposite adiabatic regime.
The conclusions are given in Sec.~\ref{sec:conclusion} while some  details of the calculations are reported in Appendices A and B.

\section{Unitary superconvergent time-dependent perturbation theory}
\label{sec:time}
\subsection{KAM algorithm for autonomous Hamiltonians}
\label{sub:KAM}
In this section, we present the KAM technique 
 for a time-independent Hamiltonian following the formulation of~\cite{Bellissard}
and using the averaging method of~\cite{SchererII}.
Let $K_{1} = K\rz + \epsilon V_{1}$ where  $K\rz$ is a reference Hamiltonian defined on a Hilbert space
$\Hs $ 
and $\epsilon V_{1}$ a bounded self-adjoint perturbation with small parameter $\epsilon$. 
As will become clear below, the subscript $n$ indicates that an operator $A_n$ is involved in the $n$-th iteration.
On the other hand, the upperscript e  stands for {\em effective} and indicates that an operator $B\rn$ constructed at the $n$-th step will be taken as the new reference at the next step.
Throughout the paper the leading order in $\ep$ will appear explicitly in front of
the operators which are thus themselves of order $\epsilon^{0}$ but may still depend on $\ep$ although we shall not indicate it explicitly. 
We look for the generator $W_{1}$ of a unitary transform
$T_{1} \equiv e^{\epsilon W_{1}}$ such that
\BEA
\label{eq:K2}
T_{1}^{\dagger}K_{1}T_{1}\equiv T_{1}^{\dagger}(K\rz + \epsilon V_{1})T_{1}
=K\ru +\epsilon^{2} V_{2}\equiv K_{2} \, ,
\EEA
with  $[K\ru,K\rz]=0$.
Writing $K\ru \equiv K\rz + \epsilon D_{1}$,
the unknown $W_{1}$ and $D_{1}$ are solutions of the following commutator equations:
\BES
\label{eq:commeq}
\BEA
& &\left[K\rz,D_{1}\right] = 0 \, , \label{eq:commeq0}\\
& &\left[K\rz,W_{1} \right] + V_{1} = D_{1} \, . \label{eq:commeq1}
\EEA
\EES
The remainder $\ep^2 V_{2}$ contains all the terms of Eq.~(\ref{eq:K2}) which are not of order $\ep^0$ (which disappear trivially) or of order $\ep$ (which disappear identically because of Eq.~(\ref{eq:commeq1})).
It reads
\BEA
\label{eq:V2bis} 
\ep^2 V_{2}&=&-\frac{\ep^2}{2} \left[W_1,V_1\right] -\frac{\ep^2}{2}\left[W_1,D_1\right]\NN \\
&+&\frac{\ep^3}{3} \left[W_1,\left[W_1,V_1\right]\right]+\frac{\ep^3}{6} \left[W_1,\left[W_1,D_1\right]\right]+\ldots , \quad 
\EEA
or, writing the series of commutator
 in a compact form
 that we shall use later,
\BE
\label{eq:V2} 
\ep^2 V_{2}=\sum_{k=1}^{\infty}\frac{(-1)^k\ep^{k+1}}{(k+1)!} \,\left\{k\, \ad^{k}(W_1,V_1) +\ad^{k}(W_1,D_1) \right\}, \
\EE
where
\BEA
\ad^{k}(A,B)\equiv \left\{
\BA{ll} 
B & k=0\\
\left[A,\ad^{k-1} (A,B) \right] \quad&  k \geq 1  .
\EA 
\right. 
\EEA
The solutions to Eqs.~(\ref{eq:commeq}) can be written in terms
of averages:
\BES
\label{eq:defD1W1}
\BEA
 D_{1}\!\!&=&\!\!\lim_{T\rightarrow \infty} \frac{1}{T}
\int_{0}^{T} e^{-itK\rz}V_{1}e^{itK\rz} \ud t \ \equiv \
 \overline{V_{1}}^{K\rz} \label{eq:defD1}  , \\
W_{1}\!\!&=&\!\!\lim_{T\rightarrow \infty}\frac{-i}{T}
\int_{0}^{T} \! \ud t
\! \int_{0}^{t} \ud t^{\prime} e^{-it^{\prime}K\rz}\left(V_{1}-\overline{V_{1}}^{K\rz}\right)
e^{it^{\prime}K\rz} \nonumber \\ 
& \equiv & - i \widehat{V_{1}}^{K\rz }. \label{eq:defW1}
\EEA
\EES
This is readily checked upon substitution, noting that $e^{-itK\rz}$ is the propagator of the reference Hamiltonian
$K\rz$ in units such that $\hbar$=1.
The shorthand notations on the right handside of Eqs. (\ref{eq:defD1W1})
can be viewed as well defined linear transformations
of the operator $V_1$.

The process can be iterated, transforming now the operator $K_2$ defined
by Eq.~(\ref{eq:K2}) with $T_2=e^{\epsilon^2 W_2}$ and considering $K\ru$
as the new reference operator:
\BEA
T_{2}^{\dagger}K_{2}T_{2}\equiv T_{2}^{\dagger}(K\ru + \epsilon^2 V_{2})T_{2}
=K\rd +\epsilon^{4} V_{3}\equiv K_{3} \, ,
\EEA
with  $[K\rd,K\ru]=0$.
Notice that the new perturbation is not of order $\ep^3$ as would be the case in a standard perturbation theory, but of order $\ep^4$.
Similarly, after $n$ iterations we obtain
\begin{equation}
T_{n}^{\dagger}K_{n}T_{n} = K\rn + \epsilon^{2^{n}} V_{n+1} \equiv K_{n+1} \, ,
\end{equation}
with $[K\rn,K\rnm]=0$.
The new reference Hamiltonian $K\rn$ is written
\BE
K\rn = K\rnm+\epsilon^{2^{n-1}} D_{n}= K\rz + \sum_{j=1}^n \epsilon^{2^{j-1}} D_{j} \, .
\EE
The generator $W_{n}$ of the unitary transformation $T_{n} = e^{\epnm W_{n}}$
and the operator $D_{n}$, solving equations analogous to Eqs.~(\ref{eq:commeq}), are  calculated as
\BES
\begin{eqnarray}
D_{n} &=&\overline{V_{n}}^{K\rnm} \, ,\\
W_{n} &=&-i\widehat{V_{n}}^{K\rnm} \, .
\end{eqnarray}
\EES
The remainder  being of order $\epsilon^{2^{n}}$, the KAM algorithm is called superconvergent.
However, we emphasize that a proof of convergence is to
be established in each case.
We also note that in the absence of resonances,
$\ep$ need not be small for this algorithm to converge~\cite{PhysicA}.

\subsection{Extended Hilbert space}
\label{sub:Kspace}
Given a time-dependent Hamiltonian $H(t)$ acting on a Hilbert space $\Hs$, we recall here how the notion of extended Hilbert space of \cite{Howland} allows   to construct a time-independent operator on that space.
Let $U_{H}(t,t_{0})$ denote the  evolution operator associated to $H(t)$,
so that the Schr\" odinger equation and the initial condition read  
\begin{equation}
\label{eq:UH}
i\ddpt U_{H}(t,t_{0}) = H(t)
U_{H}(t,t_{0}) \ , \quad  U_{H}(t_{0},t_{0}) 
= \un_{\Hs} \, ,
\end{equation}
where $\un_{\Hs}$ is the identity operator on $\Hs$. 
We introduce a parameter $s \in \mathbb {R}$ which plays the role of
an arbitrary reference time, and let $U_{H}(t,t_{0};s)$ be the
solution of Eqs.~(\ref{eq:UH}) now with $H(t+s)$.
In $\Hs$,  the operator $U_{H}(t,t_{0};s)$ depends
parametrically on $s$. 
Notice that $U_{H}(t,t_{0};0)=U_{H}(t,t_{0})$.

An extended Hilbert space $\Ks$ where $s$ is now an
additional coordinate  can be defined as the tensor product of $\Ls$ and $\Hs$ where $\Ls\equiv\mathcal{L}_2(\Rs)$ is the space of square integrable functions on the real line: $\Ks\equiv\Ls \otimes \Hs$.
The family of operators $U_{H}(t,t_{0};s)$ acting on $\Hs$ is {\em lifted} to the operator $\mathcal{U}_{\mathcal{H}}(t,t_{0};s)$ defined on $\Ks$ by considering the full dependence on $s$ as a multiplication operator on $\Ls$.
Similarly, the family of operators $H(t+s)$ on $\Hs$ is lifted to the operator $\mathcal{H}(t+s)$ on $\Ks$.
We shall denote operators acting on the extended Hilbert space $\Ks$
by uppercase calligraphic letters and shall refer
to $\mathcal{U}_{\mathcal{H}}(t,t_{0};s)$ and $\mathcal{H}(t+s)$ as the {\em lifts}
of $U_{H}(t,t_{0})$  and $H(t)$ respectively (with the understanding that the family of operators $U_{H}(t,t_{0};s)$ or $H(t+s)$ is  considered as an intermediate step).
The lift of Eqs.~(\ref{eq:UH}) on the extended Hilbert space $\Ks$ reads
\BES
\label{eq:UK}
\BEA
&& i\ddpt \mathcal{U}_{\mathcal{H}}(t,t_{0};s) = {\mathcal{H}}(t+s)\mathcal{U}_{\mathcal{H}}(t,t_{0};s) \, , \\
&& \mathcal{U}_{\mathcal{H}}(t_{0},t_{0};s) = {\un}_{\Ls} \otimes {\un}_{\Hs} \, .
\EEA
\EES
where ${\un}_{\Ls}$ is the identity operator on $\Ls$.
Notice that
\begin{eqnarray}
\mathcal{U}_{\mathcal{H}}(t+s^{\prime},t_{0}+s^{\prime};s-s^{\prime}) =\mathcal{U}_{\mathcal{H}}(t,t_{0};s) \quad \forall s^{\prime}  \in \mathbb{R} . \label{eq:prop1}
\end{eqnarray}

Finally, an  extended Hamiltonian is defined on $\Ks$ as the  time-independent self-adjoint operator ${\mathcal{K}} \equiv {\mathcal{H}}(s) -i\ddps \otimes \un_{\Hs}$.
Its associated unitary evolution operator reads
$\mathcal{U}_{\mathcal{K}}(t,t_0) \equiv e^{-i(t-t_{0}){\mathcal{K}}}$ and 
 is related to the solution of Eqs.~(\ref{eq:UK}) by the following equation which is easily derived:
\BE
\label{eq:connectKH}
\mathcal{U}_{\mathcal{K}}(t,t_{0}) ={\sf T}_{-t} \mathcal{U}_{\mathcal{H}}(t,t_{0};s) {\sf T}_{t_0} \, , 
\EE
where the translation operator ${\sf T}_t$ acts on functions $\xi(s) \in \Ls$ according to ${\sf T}_t  \xi(s)=\xi(s+t)$ and can be expressed as ${\sf T}_t=e^{t \ddps}$.

\subsection{KAM algorithm in the extended Hilbert space for non-autonomous Hamiltonians}
\label{sub:time}
We consider $\epsilon V_{1}(t)$ as a bounded time-dependent perturbation of the time-dependent reference Hamiltonian $H\rz(t)$ defined on the Hilbert space $\Hs$ and whose propagator $U_{H\rz}(t,t_0)$  is known. 
Our aim is to obtain a KAM expansion for the evolution operator $U_{H_1}(t,t_0)$ of the full Hamiltonian $H_1(t)\equiv H\rz(t) + \epsilon V_{1}(t)$.
We first consider the extended Hilbert space $\Ks=\Ls \otimes \Hs$, and the lifts  $\mathcal{U}_{\mathcal{H}_1}(t,t_0;s)$ and $\mathcal{H}_1(t+s)$ of the  operators $U_{H_1}(t,t_0)$ and $H_1(t)$ as defined in Sec.~\ref{sub:Kspace}.
Similarly, $\mathcal{U}_{\mathcal{H}\rz}(t,t_0;s)$, $\mathcal{H}\rz(t+s)$ and $\mathcal{V}_1(t+s)$ denote the lifts on $\Ks$ of $U_{H\rz}(t,t_0)$, $H\rz(t)$ and $V_{1}(t)$.
We then define the associated extended Hamiltonian on $\Ks$:

\BES
\BEA
\label{K1time}
\mathcal{K}_1 \!\!\!&\equiv&\!\!\! \mathcal{H}_1(s)  -i\ddps \otimes \un_{\Hs} , \\
\!\!\!&=&\!\!\!\mathcal{H}\rz(s)-i\ddps \otimes \un_{\Hs} + \epsilon \mathcal{V}_{1}(s) \equiv  \mathcal{K}\rz + \epsilon \mathcal{V}_{1}(s) , \qquad
\EEA
\EES
which is of the form considered in Sec.~\ref{sub:KAM}.
Hence, we can now apply the KAM technique in the extended Hilbert space to obtain a KAM expansion for the time-independent operator $\mathcal{K}_1$.

At the $n$-th iteration of the algorithm, the operator $\mathcal{K}_n=\mathcal{K}\rnm+\epnm \mathcal{V}_n(s)$ is transformed by the
unitary operator $\mathcal{T}_{n}(s)$ according to
\BEA
\label{eq:Kn+1}
\mathcal{T}_n^{\dagger}(s) \mathcal{K}_n \mathcal{T}_n(s) = \mathcal{K}\rn+\epn \mathcal{V}_{n+1}(s) \, , \ \ 
\EEA
with
\BES
 \label{TKH}
\BEA
\mathcal{K}\rn&=& \mathcal{H}\rn(s)-i\ddps \otimes \un_{\Hs}  , \label{eq:K0n}\\
\mathcal{H}\rn(s) &=& \mathcal{H}\rnm(s) + \epnm \overline{\mathcal{V}_{n}}^{\mathcal{K}\rnm}(s)  , \label{eq:H0n}\\
\mathcal{T}_n(s)&=& \exp\left(-i\epnm \widehat{\mathcal{V}_{n}}^{\mathcal{K}\rnm}(s)\right)  .\label{eq:Tn} 
\EEA
\EES
The remainder $\epn \mathcal{V}_{n+1}(s)$ is given by an expression analogous to Eq.~(\ref{eq:V2}):
\begin{widetext}
\BEA
\label{eq:Vn+1}
\epn \mathcal{V}_{n+1}(s)=\sum_{k=1}^{\infty}\frac{i^k\ep^{(k+1)2^{n-1}}}{(k+1)!} \left\{k\, \ad^{k}\left(\widehat{\mathcal{V}_{n}}^{\mathcal{K}\rnm}(s),\mathcal{V}_n(s)\right)  + \ad^{k}\left(\widehat{\mathcal{V}_{n}}^{\mathcal{K}\rnm}(s),\overline{\mathcal{V}_{n}}^{\mathcal{K}\rnm}(s)\right)\right\}  . \quad
\EEA
The operators $\overline{\mathcal{V}_{n}}^{\mathcal{K}\rnm}(s)$ and
$\widehat{\mathcal{V}_{n}}^{\mathcal{K}\rnm}(s)$, defined by Eqs.~(\ref{eq:defD1W1}),  
can be expressed in terms of the operator $\mathcal{U}_{\mathcal{H}\rnm}(t,t_0;s)$ using Eq.~(\ref{eq:connectKH}) for the propagator of $\mathcal{K}\rn$:
%\begin{widetext}
\BES
\label{eq:Vbhtime}
\BEA
\label{eq:Vbtime}
\overline{\mathcal{V}_{n}}^{\mathcal{K}\rnm}(s)\!\!&=&\!\!\lim_{T\rightarrow \infty} \frac{1}{T}
\int_{0}^{T} \! \ud t \, \mathcal{U}_{\mathcal{H}\rnm}(0,-t;s) 
\mathcal{V}_{n}(s-t)\mathcal{U}_{\mathcal{H}\rnm}^{\dagger}(0,-t;s)\equiv \overline{\mathcal{V}_{n}}^{\mathcal{H}\rnm}(s) ,\\
\widehat{\mathcal{V}_{n}}^{\mathcal{K}\rnm}(s)\!\!&=&\!\!  \lim_{T\rightarrow \infty} \frac{1}{T} 
\int_{0}^{T} \! \ud t^{\prime}\! \int_{0}^{t^{\prime}} \! \ud t \, \mathcal{U}_{\mathcal{H}\rnm}(0,-t;s)
\left[\mathcal{V}_{n}(s-t)-\overline{\mathcal{V}_{n}}^{\mathcal{K}\rnm}(s-t)\right]
\mathcal{U}_{\mathcal{H}\rnm}^{\dagger}(0,-t;s)\equiv \widehat{\mathcal{V}_{n}}^{\mathcal{H}\rnm}(s)  . \qquad 
\label{eq:Vhtime}
\EEA
\EES
On the other hand, from Eqs.~(\ref{eq:connectKH}) and (\ref{TKH}) and the fact that $\mathcal{K}\rn$ commutes with all the operators $\mathcal{K}\rk$ constructed at the preceding
iterations one deduces that
\BES
\label{eq:UH0n}
\BEA
 \mathcal{U}_{\mathcal{H}\rn}(t,t_0;s)\!&=&\! \mathcal{U}_{\mathcal{H}\rz}(t,t_0;s) \,\mathcal{S}_{1}(t,t_0;s+t_0)  \ldots \mathcal{S}_{n}(t,t_0;s+t_0) , \qquad \label{eq:UH0na}\\
 \!&=&\!\mathcal{S}_{1}(t,t_0;s+t) \ldots \mathcal{S}_{n}(t,t_0;s+t) \,\mathcal{U}_{\mathcal{H}\rz}(t,t_0;s)  , \qquad \label{eq:UH0nb}
\EEA
\EES
\end{widetext}
where $\mathcal{S}_p(t,t_0;s)$ denotes the following operator on $\Ks$:
\BEA
\label{eq:S}
\mathcal{S}_p(t,t_0;s)\equiv\exp{\left(-i (t-t_0) \eppm \overline{\mathcal{V}_{p}}^{\mathcal{H}\rpm}(s) \right)} .
\EEA
The detailled derivation of Eqs.~(\ref{eq:Vbhtime}) and (\ref{eq:UH0n}) is provided
in appendix \ref{ap:time}. 
It follows that
$\overline{\mathcal{V}_{n}}^{\mathcal{H}\rnm}(s)$ and  $\widehat{\mathcal{V}_{n}}^{\mathcal{H}\rnm}(s)$  are calculated from $\mathcal{U}_{\mathcal{H}\rz}(t,t_0;s)$ as well as the operators $\overline{\mathcal{V}_{p}}^{\mathcal{H}\rpm}(s)$ and  $\widehat{\mathcal{V}_{p}}^{\mathcal{H}\rpm}(s)$  constructed at the preceding iterations.
Hence, the operators $\mathcal{T}_n(s)$, $\mathcal{K}\rn$ and $\mathcal{V}_{n+1}(s)$ entering Eq.~(\ref{eq:Kn+1}) are now entirely determined.

The extended Hamiltonian $\mathcal{K}_1$ we started with can be expressed in terms of $\mathcal{K}\rn$ by repeated use of Eq.~(\ref{eq:Kn+1}):
\BE
\label{K1Tn}
\mathcal{K}_1=\mathcal{T}_{1}(s)\ldots \mathcal{T}_{n}(s)\,\mathcal{K}\rn\, \mathcal{T}_{n}^{\dagger}(s) \ldots \mathcal{T}_{1}^{\dagger}(s) + \mathcal{O}(\epn) .
\EE
The propagator $\mathcal{U}_{\mathcal{K}_1}(t,t_0)=e^{-i(t-t_0)\mathcal{K}_1}$ allows then to construct 
the operator $\mathcal{U}_{\mathcal{H}_1}(t,t_{0};s)$ on $\Ks$ from Eq.~(\ref{eq:connectKH}).
Taking also Eq.~(\ref{eq:UH0na}) into account yields
\BEA
\mathcal{U}_{\mathcal{H}_1}(t,t_{0};s)=\!\!\!&&\!\!\!\mathcal{T}_{1}(t+s) \ldots\mathcal{T}_{n}(t+s) \,\mathcal{U}_{\mathcal{H}\rz}(t,t_{0};s) \NN\\\!\!\!&&\!\!\!\mathcal{S}_{1}(t,t_0;s+t_0) \ldots\mathcal{S}_{n}(t,t_0;s+t_0) \NN\\
\!\!\!&&\!\!\!\mathcal{T}_{n}^{\dagger}(t_0+s) \ldots \mathcal{T}_{1}^{\dagger}(t_0+s) + \mathcal{O}(\epn)  . \quad \label{UH}
\EEA

It is now possible to return to the original Hilbert space $\Hs$,  by considering the dependence on the $s-$variable of each of the operators entering Eq.~(\ref{UH}), which defines a multiplication operator on $\Ls$, as a parametric dependence on time in  $\Hs$, and subsequently setting $s=0$. 
Hence, in agreement with our notations, Eq.~(\ref{UH}) is the lift on $\Ks$ of the following expression  for the propagator $U_{H_1}(t,t_{0}) $ on $\Hs$:
\BEA
U_{H_1}(t,t_{0})=\!\!\!&&\!\!\!T_{1}(t) \ldots T_{n}(t)\, U_{H\rz}(t,t_{0}) S_{1}(t,t_{0};t_0) \ldots \NN\\ 
\!\!\!&&\!\!\! S_{n}(t,t_{0};t_0)\,T_{n}^{\dagger}(t_0)\ldots  T_{1}^{\dagger}(t_0)+\mathcal{O}(\epn) , \qquad \label{eq:calUH}
\EEA
where $S_{p}(t,t_0;t_0)$ and $T_{p}(t)$ acting on $\Hs$ are obtained, as just described, from their lift $\mathcal{S}_{p}(t,t_0;s+t_0)$ and  $\mathcal{T}_{p}(t+s)$ constructed on $\Ks$.

In practice, however,  we shall find it simpler to construct $S_{p}(t,t_0;t_0)$ and $T_{p}(t)$ directly in $\Hs$.
As we show  below, this can be achieved by considering Eqs.~(\ref{eq:H0n})-(\ref{eq:S}) which have well defined meaning on $\Ks$ as the lift of equations for  corresponding operators defined on $\Hs$.
In particular, Eq.~(\ref{eq:Vbtime}) with $n=1$ is the lift of a similar equation defining the operator $\overline{V_{1}}^{H\rz}(t)$  on $\Hs$ in terms of $U_{H\rz}(t,t_0)$ and $V_1(t)$.

\subsection{KAM expansion in the original Hilbert space for non-autonomous evolution operators}
\label{sub:original}
In this section, we shall construct  the propagator $U_{H_1}(t,t_0)$ from  Eq.~(\ref{eq:calUH}) through an iterative procedure entirely defined in the Hilbert space $\Hs$ of the Hamiltonian $H_1(t)=H\rz(t)+\ep V_1(t)$, i.e. we shall not have to define operators in an extended Hilbert space.
The operators $U_{H\rz}(t,t_0)$ and $V_1(t)$ allow to construct  the operator $\overline{V_{1}}^{H\rz}(t)$ on $\Hs$ according to Eq.~(\ref{eq:calVbtime}) given below, where we set $p=1$.
Subsequently, the operator $\widehat{V_{1}}^{H\rz}(t)$  can be obtained  from  these operators using Eq.~(\ref{eq:calVhtime}) with $p=1$.
Hence, by Eqs.~(\ref{eq:calSTH}), the operators $S_1(t,t_0;t_0)$ and $T_1(t)$ are determined.

On the other hand, Eq.~(\ref{eq:calVn+1}) with $p=1$ enables us to derive the operator $V_2(t)$ on $\Hs$ from the operators  $\overline{V_{1}}^{H\rz}(t)$ and $\widehat{V_{1}}^{H\rz}(t)$ we have just constructed. 
Similarly, Eq.~(\ref{eq:calUH0na}) yields the propagator $U_{H\ru}(t,t_0)$.
It follows that the operators $\overline{V_{2}}^{H\ru}(t)$ and $\widehat{V_{2}}^{H\ru}(t)$  can be obtained from Eqs.~(\ref{eq:calVbtime}) and (\ref{eq:calVhtime}) now with $p=2$.

For $p\geq 1$ we have the following operators on $\Hs$:
\BES
\label{eq:calSTH}
\BEA
\label{eq:calS}
S_p(t,t_0;t_1)\!\!&= &\!\!\exp{\left(-i (t-t_0) \eppm \overline{V_p}^{H\rpm}(t_1) \right)} ,\qquad\\
T_p(t)\!\!&= &\!\!\exp{\left(-i \eppm \widehat{V_p}^{H\rpm}(t) \right)}, \qquad \label{eq:calT}
\EEA
\EES
where
\begin{widetext}
\BES
\label{eq:calVbhtime}
\BEA
\label{eq:calVbtime}
\overline{V_{p}}^{H\rpm}(t)\!\!&=&\!\!\lim_{T\rightarrow \infty} \frac{1}{T}
\int_{t-T}^{t} \! \ud u \, U_{H\rpm}(t,u) 
V_{p}(u)U_{H\rpm}^{\dagger}(t,u), \quad \\
\widehat{V_{p}}^{H\rpm}(t)\!\!&=&\!\!  \lim_{T\rightarrow \infty} \frac{1}{T} 
\int_{0}^{T} \! \ud t^{\prime}\! \int_{t-t^{\prime}}^{t} \! \ud u \, U_{H\rpm}(t,u)
\left[V_{p}(u)-\overline{V_{p}}^{H\rpm}(u)\right]
U_{H\rpm}^{\dagger}(t,u)  , 
\label{eq:calVhtime}\\
\label{eq:calVn+1}
V_{p+1}(t)\!\!&=&\!\!\sum_{k=1}^{\infty}
\frac{i^{k}\ep^{(k-1)2^{p-1}}}{(k+1)!}\, \left\{ k\, \ad^{k}
\left(\widehat{V_p}^{H\rpm}(t),V_p(t)\right)
 + \ad^{k}\left(\widehat{V_p}^{H\rpm}(t),\overline{V_p}^{H\rpm}(t)\right)\right\} , \quad
\EEA
\EES
\end{widetext}
and
\BEA
U_{H\rp}(t,t_0)\!&=&\! U_{H\rz}(t,t_0) \, S_{1}(t,t_0;t_0) \ldots S_{p}(t,t_0;t_0) . \qquad \label{eq:calUH0na}
\EEA
Note that
$H\rp(t)=H\rpm(t) + \eppm \overline{V_{p}}^{H\rpm}(t)$. 
By the iterative procedure described here and which rests solely on Eqs.~(\ref{eq:calSTH})-(\ref{eq:calUH0na}), the operators $S_{p}(t,t_0;t_0)$ and $T_p(t)$ are constructed entirely in the Hilbert space $\Hs$.
The propagator $U_{H_1}(t,t_{0})$ of the non-autonomous  Hamiltonian $H_1(t)=H\rz(t)+\ep V_1(t)$ is then obtained by Eq.~(\ref{eq:calUH}) up to a desired order in $\ep$.

\section{Pulse-driven systems}
\label{sec:pulse}
\subsection{General case}
\label{sub:general}
In this section, we consider the physically relevant case of time-dependent perturbations $\ep V_1(t)$ which are switched on at a given finite time $\ti$. 
To allow for some flexibility in the choice of the reference operators we shall consider the slightly more general case of perturbations $\ep V_1(t)$  which before $\ti$ are constant in time and commute with the reference propagator $U_{H\rz}(t,t_0)$ on $\Hs$:
\BES
\label{eq:VUti}
\BEA
&&V_1(t)=V_1(\ti) \quad \forall t \leq \ti ,\label{eq:Vti} \\
&&\left[V_1(\ti),U_{H\rz}(t^{\prime},t)\right]=0 \quad \forall t , t^{\prime} \leq \ti . \label{eq:Uti}
\EEA
\EES
After $\ti$ the time-dependence of $V_1(t)$ is supposed to be uniformly bounded in time but ortherwise arbitrary.
In particular, it need not be turned on or off infinitely slowly or rapidly, and need not be constant or periodic in the meantime.
For this class of perturbations, that we refer to as {\em pulsed perturbations},  the limits in Eqs.~(\ref{eq:calVbhtime}) can be calculated as we show in appendix~\ref{ap:pulse}.

On the one hand, from Eq.~(\ref{eq:calVbtime}) we obtain
\BES
\label{eq:Vbti}
\BEA
\label{eq:Vb1ti}
&&\!\!\overline{V_{1}}^{H\rz}(t)=U_{H\rz}(t,\ti) 
V_{1}(\ti)U_{H\rz}^{\dagger}(t,\ti) , \qquad \\
&&\!\!\overline{V_{p}}^{H\rpm}(t)=0 \quad \forall  p > 1 . \quad \label{eq:Vbnti}
\EEA
\EES\
Hence,   Eq.~(\ref{eq:calS}) yields
\BES
\label{eq:calSn}
\BEA
\label{eq:calS1}
S_{1}(t,t_{0};t_0)\!\!\!&=&\!\!\!U_{H\rz}(t_{0},\ti)e^{-i(t-t_0)\ep V_1(\ti)}U^{\dagger}_{H\rz}(t_{0},\ti) , \qquad \\
\label{eq:calS2}
S_{p}(t,t_{0};t_0)\!\!\!&=&\!\!\!\un_{\Hs}  \quad \forall p > 1 , 
\EEA
\EES
which by Eq.~(\ref{eq:calUH0na}) implies
\BEA
\label{eq:calUH0nati}
U_{H\ru}(t,t_0) =U_{H\rz}(t,\ti)e^{-i(t-t_0)\ep V_1(\ti)}U_{H\rz}^{\dagger}(t_0,\ti).
\EEA
Note that $H\rpm(t)=H\ru(t)$ for all  $p > 1$.

On the other hand, Eq.~(\ref{eq:calVhtime}) results in
%\begin{widetext}
\BES
\label{eq:Vhti}
\BEA
\!\!\!&&\!\!\!\widehat{V_1}^{H\rz}(t)=\int_{\ti}^{t} \! \ud u   U_{H\rz}(t,u) \left[V_1(u) -\overline{V_1}^{H\rz}(u)\right] U_{H\rz}^{\dagger}(t,u) , \NN\\
\!\!\!&&\!\!\!\\
\!\!\!&&\!\!\!\widehat{V_{p}}^{H\rpm}(t)=
\int_{\ti}^{t} \! \ud u \, U_{H\ru}(t,u) V_{p}(u) U_{H\ru}^{\dagger}(t,u)\quad  \forall p > 1,\NN\\ 
\!\!\!&&\!\!\!\label{eq:Vnhti}
\EEA
\EES
%\end{widetext}
where $V_p(t)$ is given by Eq.~(\ref{eq:calVn+1}).  

For pulsed perturbations it follows that  Eq.~(\ref{eq:calUH}) for the propagator $U_{H_1}(t,t_0)$ up to correction terms of order $\epn$ becomes

\BEA
U_{H_1}(t,t_{0})=\!\!\!&&\!\!\!T_{1}(t) \ldots T_{n}(t)\,  U_{H\rz}(t,\ti)e^{-i(t-t_0)\ep V_1(\ti)}\NN\\
\!\!\!&&\!\!\!U^{\dagger}_{H\rz}(t_{0},\ti)\,T_{n}^{\dagger}(t_0)\ldots  T_{1}^{\dagger}(t_0)+\mathcal{O}(\epn) , \qquad \label{eq:calUHpulse}
\EEA
where $T_p(t)$ defined by Eq.~(\ref{eq:calT}) is obtained through the simple integral given in Eqs.~(\ref{eq:Vhti}).

\subsection{Exact resummation of the remainder $\epp V_{p+1}(t)$ for pulse-driven two-level systems}
\label{sub:resum}
For two-level systems some of the formulas that we have constructed
in the preceding sections can be written in an explicit simple form.
In particular, we shall calculate exactly the remainders of the KAM iterations.
The partition of an Hamiltonian $H_1(t)=H\rz(t)+\ep V_1(t)$ on $\Hs=\mathbb{C}^2$ can always be choosen such that
\BE
\label{eq:V12level}
V_1(t)=\sum_{k=1}^3 v_k(t)\sigma_k ,
\EE
where $v_k(t) \in \mathcal{L}_2(\mathbb{R})$ are real functions on
$\mathbb{R}$ and $\sigma_k$ the  Pauli matrices:
\BE
\sigma_1=\begin{pmatrix}
0 & 1 \\ 1 & 0
\end{pmatrix},\quad \sigma_2=\begin{pmatrix}
0 & -i \\ i & 0
\end{pmatrix}, \quad \sigma_3=\begin{pmatrix}
1 & 0 \\ 0 & -1
\end{pmatrix}.
\EE 
The perturbation $\ep V_1(t)$ is switched on at a finite time $\ti$ so that Eqs.~(\ref{eq:Vbti}) reduce to $\overline{V_p}^{H\rz}(t)=0$ for all $p\geq 1$, and  Eqs.~(\ref{eq:Vhti}) to
\BEA
\label{eq:Vh2level}
\widehat{V_p}^{H\rz}(t)=\int_{\ti}^{t} \! \ud u \,U_{H\rz}(t,u) V_p(u) U_{H\rz}^{\dagger}(t,u) \quad  \forall p \geq 1. \qquad
\EEA
Furthermore, the infinite series of Eq.~(\ref{eq:calVn+1}) for 
$V_{p+1}(t)$ with $p \geq 1$ reads 
\BE
\label{eq:Vk+12level}
V_{p+1}(t)=\sum_{k=1}^{\infty}
\frac{k \, i^{k} \,\ep^{(k-1)2^{p-1}}}{(k+1)!} \, \ad^{k}
\left(\widehat{V_{p}}^{H\rz}(t),V_p(t)\right).
\EE 

Let $B_p(t)$ be the unitary matrix  which diagonalizes $\widehat{V_{p}}^{H\rz}(t)$.
As the matrix $V_{1}(t)$ is traceless, it is straightforward to show by induction with the help of Eqs.~(\ref{eq:Vh2level}) and (\ref{eq:Vk+12level}) that $\widehat{V_{p}}^{H\rz}(t)$ is traceless for all $p\geq 1$. Hence
\BEA
\label{eq:Vbdiag}
B_p^{\dagger}(t) \widehat{V_{p}}^{H\rz} (t) B_p(t) = - r_p(t)\sigma_3,
\EEA
where  
\BE
r_p(t)\equiv [-\det \widehat{V_{p}}^{H\rz}(t)]^{1/2} ,
\EE
is real.
The following identity holds for all $k \geq 1$:
\BEA
\label{eq:adVVb}
\ad^{k}\left(\widehat{V_{p}}^{H\rz},V_p\right)
=B_p \ad^{k}\left(-r_p \sigma_3,B_p^{\dagger}V_pB_p\right) B_p^{\dagger},
\EEA
and for any Hermitian matrix $M$ in $\mathbb{C}^2$ one has
\BEA
\label{eq:ad3VVb}
\ad^{3}(\sigma_3,M)=4 \ad^{1}(\sigma_3,M) .\quad
\EEA
Combining Eqs.~(\ref{eq:adVVb}) and (\ref{eq:ad3VVb}) yields
\BEA
\ad^{k}\left(\widehat{V_{p}}^{H\rz},V_p\right)=
(-2  r_p)^{k-\ell} \ad^{\ell}\left(\widehat{V_{p}}^{H\rz},V_p\right), 
\EEA
where $\ell$ is 1 if $k$ is odd, and 2 if $k$ is even.
The series of Eq.~(\ref{eq:Vk+12level}) for $V_{p+1}(t)$ can then be cast into the form
\begin{widetext}
\BEA
\label{eq:resum}
V_{p+1}(t) &=& \xi_p(t)  \left[\widehat{V_{p}}^{H\rz}(t),V_p(t)\right] 
+\eppm\gamma_p(t) \left[\widehat{V_{p}}^{H\rz}(t),\left[\widehat{V_{p}}^{H\rz}(t),V_p(t)\right]\right],
\EEA
where in agreement with our notations the following quantity are of order $\ep^0$:
\BES
\label{xigamma}
\BEA
\xi_p(t)&=&i\frac{\cos[2 \eppm r_p(t)]-1 +2  \eppm r_p(t) \sin[2 \eppm r_p(t)]}{[2 \eppm r_p(t)]^2} ,\\
\gamma_p(t)&=&\frac{2  \eppm r_p(t) \cos[2 \eppm r_p(t)]-\sin[2 \eppm r_p(t)]}{[2\eppm r_p(t)]^3 }.
\EEA
\EES
\end{widetext}
The remainder $\epp V_{p+1}(t)$  is therefore well-defined for all values of the parameter $\ep$ and all times $t$.
We then obtain from Eq.~(\ref{eq:calUHpulse}) the propagator $U_{H_1}(t,t_0)$ up to an arbitrary order in $\ep$, 
\BEA
U_{H_1}(t,t_{0})=U_{H_1}^{(n)}(t,t_{0})+ \mathcal{O}(\epn), \label{eq:calUH2}
\EEA
where 
\BE
U_{H_1}^{(n)}(t,t_{0})\equiv T_{1}(t) \ldots T_{n}(t) U_{H\rz}(t,t_{0}) T_{n}^{\dagger}(t_0) \ldots T_{1}^{\dagger}(t_0). \quad \label{eq:calUHn}
\EE
Moreover, using Eqs.~(\ref{eq:calT}) and (\ref{eq:Vbdiag}) one deduces that
\BE
\label{eq:calTk2}
T_p(t)=\cos[\eppm r_p(t)] \un_{\Cs} -i \frac{ \sin[\eppm r_p(t)]}{r_p(t)} \widehat{V_{p}}^{H\rz}(t)  , \quad
\EE
where $\widehat{V_{p}}^{H\rz}(t)$ is given by Eq.~(\ref{eq:Vh2level}), and owing to Eq.~(\ref{eq:resum}) can be calculated for arbitrary $p$ without having to resort to an infinite series.
Note that the form of Eqs.~(\ref{xigamma}) suggests that Eqs.~(\ref{eq:resum}) and (\ref{eq:calTk2}) remain valid for finite values of the parameter $\ep$, possibly larger than unity. 
We shall see below that this is indeed what is observed numerically.

\subsection{Numerical implementation for a two-level system}
\label{sub:results}
Here we investigate the convergence of the KAM technique
with the number of iterations as well as its domain of validity  for a specific two-level model perturbed by a pulsed interaction.
The algorithm is implemented numerically for a system described by the time-independent  Hamiltonian $\omega \sigma_3$ and which interacts through $\sigma_1$ with a sine-squared pulse of characteristic duration $\tau$.
This pulse shape is commonly used
in the literature because of its bounded support and continuous
first derivative at the boundaries.
Defining the characteristic duration $\tau$ as twice the full width at half maximum fixes the total duration of a cycle to $\tau$ and yields the following dimensionless pulse shape between the dimensionless time $t_i=0$ and $t_f=1$:
\BE
\Omega(t)=\left\{\begin{array}{cc}2 A\sin^2\left(\pi t\right)\quad &
 0\le t\le 1,\\
0 & \text{ elsewhere}.\end{array}\right.
\EE
Note that the peak amplitude is twice the pulse area $A\equiv A(\tf)$ where $A(t)\equiv\int_{\ti}^t \Omega(u) \, \ud u$,  and that it can be fixed independently of the parameter $\ep \equiv \omega \tau$ that we shall take here as the small parameter.
This allows, in particular,  to treat large non-perturbative areas for short pulse durations, which corresponds to the experimental  conditions used to generate short laser pulses.
The Schr\"odinger equation reads
\begin{equation}
\label{eq:Us}
i\ddpt U(t,t_{0}) = \left[\Omega(t)\sigma_1+\epsilon \sigma_3 \right]
U(t,t_{0}),   \quad
\end{equation}
with $U(t_0,t_{0}) =\un_{\Cs}$.
For  $\epsilon=0$ its solution  is
\BE
\label{eq:U0}
U^{(0)}(t,t_{0}) \equiv e^{-i[A(t)-A(t_0)]\sigma_1} .
\EE

As a first step, we write 
Eq.~(\ref{eq:Us})  in the interaction
representation with the help of the unitary operator $U^{(0)}(t,\ti)$: 
\BE
\label{eq:Uinter}
i\ddpt U_{H_1}(t,t_{0}) = H_1(t) U_{H_1}(t,t_{0}),
\EE
where
\BES
\label{eq:HU}
\BEA
H_1(t) & \equiv & \epsilon {U^{(0)}}^{\dagger}(t,\ti)\sigma_3
U^{(0)}(t,\ti),  \label{defH1}\\
 U_{H_1}(t,t_0) & \equiv & {U^{(0)}}^{\dagger}(t ,\ti) U(t,t_0)
U^{(0)}(t_0,\ti) \label{defUH1} .
\EEA
\EES
Note that $H_1(\ti)=\ep \sigma_3$ so that if we substract $\ep \sigma_3$ from Eq.~(\ref{defH1}) we obtain an operator that vanishes for $t=\ti$. 
It is interesting  to identify the latter as the perturbation $\ep V_1(t)$ 
in order to get rid of the average $\overline{V_1}^{H\rz}(t)$ [cf Eq.~(\ref{eq:Vb1ti})] as was done in Sec.~\ref{sub:resum}:
\BEA
\label{eq:V1soudain}
V_1(t)&\equiv&  {U^{(0)}}^{\dagger}(t,\ti)\sigma_3 U^{(0)}(t,\ti) - \sigma_3, \NN \\
&=& \left(\cos [2A(t)] -1\right)\sigma_3+ \sin[ 2A(t)] \sigma_2.
\EEA
The reference operators are thus
\BES
\label{eq:HUrz}
\BEA
H\rz &=&\ep \sigma_3, \\
U_{H\rz}(t,t_0)&=&e^{-i(t-t_0) \ep \sigma_3}.
\EEA
\EES

The KAM algorithm is now applied to Eq.~(\ref{eq:Uinter}).
After $n$  iterations this results in Eq.~(\ref{eq:calUH2}).
The propagator $U(t,t_0)$ is then obtained from Eq.~(\ref{defUH1}): 
\BEA
U(t,t_{0})=U^{(n)}(t,t_{0})+\mathcal{O}(\epn) ,
\EEA
where 
\BEA
U^{(n)}(t,t_{0})\equiv U^{(0)}(t,\ti) U_{H_1}^{(n)}(t,t_0)
{U^{(0)}}^{\dagger}(t_0 ,\ti), \label{Usoudain}
\EEA
in which $U_{H_1}^{(n)}(t,t_0)$ is given by Eq.~(\ref{eq:calUHn}).

Let $|+1\rangle$ and $|-1\rangle$ denote the first and second column of $\sigma_3$ respectively.
For given $\ep$ and $A$, the wave function at the time the pulse is switched on is $\psi(\ti)= |-1\rangle$.
At the end of the pulse, the error between the
wave function $\psi(\tf)=U(\tf,\ti) \psi(\ti)$ computed by solving numerically the Schr\"odinger
equation and the wave function $\psi^{(n)}(\tf)=U^{(n)}(\tf,\ti) \psi(\ti)$ obtained after $n$ KAM iterations
(where the integration in Eq.~(\ref{eq:Vh2level}) is performed numerically)  is defined as
\BEA
\label{Err}
\Delta_n\equiv\left[\sum_{\eta=\pm 1} \left|\langle \eta |\psi(\tf) \rangle- \langle \eta |\psi^{(n)}(\tf) \rangle\right|^2 \right]^{1/2}. \quad
\EEA

\begin{figure}[h]
\includegraphics[scale=0.7]{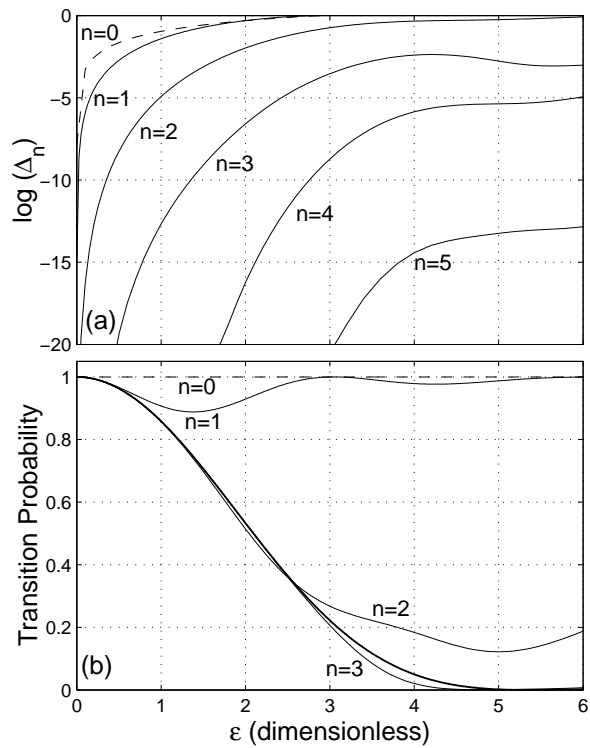}
\caption{\label{Err_epsi} For $A=\frac{\pi}{2}$ and different number of KAM iterations
$n$, (a) natural logarithm of the error $\Delta_n$ defined in 
Eq.~(\ref{Err}), and
(b) transition probabilities $|\langle+1|\psi\rangle|^2$
  computed
numerically  (thick line) and $|\langle+1|\psi^{(n)}\rangle|^2$  calculated with the KAM algorithm (thin lines)
as a function of $\epsilon$, at the end of the pulse.}
\end{figure}
Figure \ref{Err_epsi}a displays the accuracy of the KAM algorithm for
different number of iterations as a function of $\epsilon$ in the case  where the pulse area $A=\frac{\pi}{2}$, which corresponds to a peak amplitude equal to $\pi$.
One can see that this
procedure reproduces the numerical results with great accuracy for
{\it  any value of $\epsilon$} provided a sufficient (yet small) number of iterations
is used. 
For the range of $\epsilon$ shown on Fig.~\ref{Err_epsi}, the fifth iteration
is indistinguishable from the numerical result. 
For $\epsilon=0$, the peak amplitude considered here leads to a complete population transfer from the lower to the upper state (the so-called
``$\pi-$ pulse'' transfer). 
Figure~\ref{Err_epsi}b shows that this transfer decreases
for larger $\epsilon$ until becoming negligible beyond $\epsilon\approx5$,
a feature which characterizes the adiabatic regime.
It is striking that the adiabatic regime can be reached with great accuracy
from the third iteration on.
\begin{figure}[h]
\includegraphics[scale=0.7]{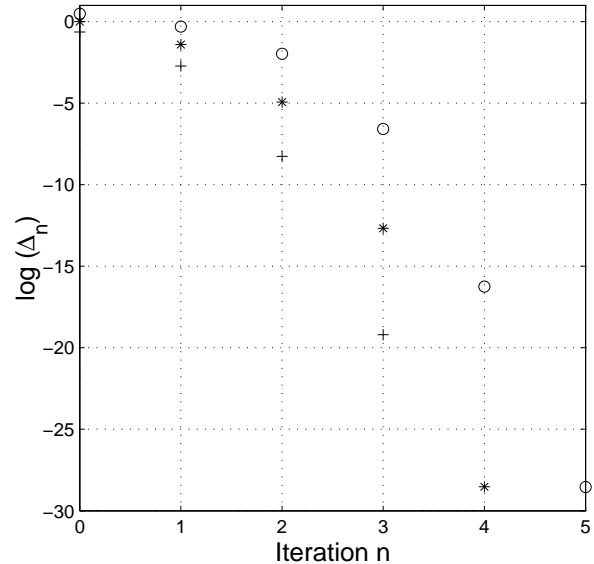}
\caption{\label{Err_it} Natural logarithm of the error $\Delta_n$ as a function
of the number $n$ of KAM iterations for $A=\frac{\pi}{2}$ and $\epsilon=0.5$ (crosses), 1 (stars) and 2 (circles).}
\end{figure}

The accuracy of the KAM algorithm
is plotted as a function of the number of iterations in Figure \ref{Err_it}.
As suggested by the order $\epsilon^{2^{n}}$ of the remainder, the error decreases faster than exponentially.

\begin{figure}[h]
\includegraphics[scale=0.7]{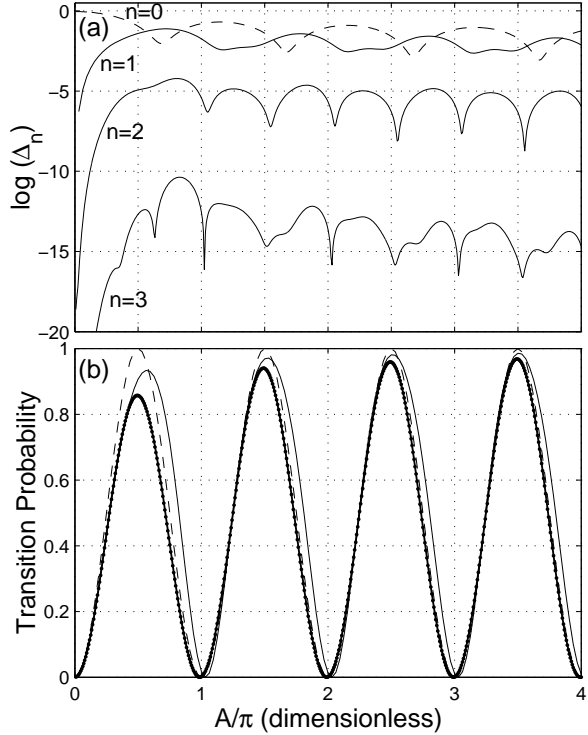}
\caption{\label{Err_aire} For $\epsilon=1$ and different KAM iterations $n$,
(a) natural logarithm of $\Delta_n$, and (b) transition probabilities
computed numerically (thick line) and with the KAM algorithm ($n=0$: dashed
 line, $n=1$: thin full line, $n=2$: dotted line, superimposed on the
numerical solution) as a function of $A/\pi$, at the end of the pulse.}
\end{figure}

Figure \ref{Err_aire} displays the accuracy of the KAM algorithm
as a function of the pulse area $A$. As expected by inspection
of the Schr\"odinger equation (\ref{eq:Us}),
Fig. \ref{Err_aire}b shows that for larger pulse area
the pulse is effectively more sudden, since
the transition probability can reach maximum values closer
to 1. The KAM algorithm accuracy is consequently globally better,
except for pulse area smaller than $\pi/2$, as seen on
Fig. \ref{Err_aire}a.

\section{Conclusion}
\label{sec:conclusion}
We have derived a unitary superconvergent algorithm, based on the KAM technique,
that allows to treat time-dependent perturbations that are localized in time.
In the physically relevant case of perturbations that are switched on at some finite time in the past, we have shown that the computation of the KAM transformations can be greatly simplified.
The remarkable efficiency of the method has been shown for a pulse-driven
two-level system, for which we obtain convergence all the way from the sudden regime to the opposite adiabatic regime.
We anticipate interesting applications of this method
in the context of alignment and orientation of molecules by pulsed
laser fields.

%\eject

\begin{acknowledgments}
This research was financially supported in part by the {\it Action Concert\'{e}e
Incitative Photonique} from the French Ministry of Research, the
{\it Conseil R\'{e}gional de Bourgogne} and a CGRI-FNRS-CNRS cooperation. D.~D. is grateful to G. Nicolis for stimulating discussions and acknowledges financial support from the Belgian FNRS.
\end{acknowledgments}

\eject

\appendix

\section{KAM algorithm in the extended Hilbert space}
\label{ap:time}
\subsection{$\overline{\mathcal{V}}^{\mathcal{K}}(s)$ and $\widehat{\mathcal{V}}^{\mathcal{K}}(s)$}
\label{sub:VW}
For time-dependent problems, the KAM algorithm involves calculating the following transforms of operators $\mathcal{V}(s)$ with respect to the propagator of  $\mathcal{K}=\mathcal{H}(s)-i\ddps \otimes \un_{\Hs}$ on the extended Hilbert space $\Ks=\Ls \otimes \Hs$:
\label{Vbh}
\BES
\BEA
\overline{\mathcal{V}}^{\mathcal{K}}(s) \!\!&\equiv&\!\!\lim_{T\rightarrow \infty} \frac{1}{T}
\int_{0}^{T} \!\!\! \ud t \, e^{-it\mathcal{K}}\mathcal{V}(s)
e^{it\mathcal{K}} \, , \label{Vb}\\
\widehat{\mathcal{V}}^{\mathcal{K}}(s)\!\!&\equiv&\!\!\lim_{T\rightarrow \infty} \frac{1}{T}
\int_{0}^{T} \!\!\! \ud t^{\prime}
\! \int_{0}^{t^{\prime}} \!\!\! \ud t  \,e^{-it\mathcal{K}}
\left(\mathcal{V}(s)-\overline{\mathcal{V}}^{\mathcal{K}}(s)\right)
e^{it\mathcal{K}}  . \NN\\
\label{Vh}
\EEA
\EES
Hence, one has to consider operators on $\Ks$
of the form $\mathcal{B}(s,t) \equiv  e^{-it\mathcal{K}} \mathcal{A}(s)e^{it\mathcal{K}}$ with $\mathcal{A}\equiv \mathcal{V}$ for $\overline{\mathcal{V}}^{\mathcal{K}}$ and $\mathcal{A} \equiv \mathcal{V}-\overline{\mathcal{V}}^{\mathcal{K}}$ for $\widehat{\mathcal{V}}^{\mathcal{K}}$. 
Using Eq~(\ref{eq:connectKH}), $\mathcal{B}(s,t)$ becomes
\begin{eqnarray}
\mathcal{B}(s,t) \!&=&\! {\sf T}_{-t}\mathcal{U}_\mathcal{H}(t,0;s)\mathcal{A}(s)\mathcal{U}_\mathcal{H}^{\dagger}(t,0;s){\sf T}_t \nonumber \\
&=&\! \mathcal{U}_\mathcal{H}(t,0;s-t)\mathcal{A}(s-t)\mathcal{U}_\mathcal{H}^{\dagger}(t,0;s-t) \nonumber \\
&=&\! \mathcal{U}_\mathcal{H}(0,-t;s)\mathcal{A}(s-t)\mathcal{U}_\mathcal{H}^{\dagger}(0,-t;s) \, , \ \
%&=& U_{\mathcal{H}}(s,s-t)\mathcal{A}(s-t)
%U_{\mathcal{H}}^{\dagger}(s,s-t),
\label{Bst}
\end{eqnarray}
where we used the definition of the translation operator and Eq.~(\ref{eq:prop1})
to obtain the second and third equalities, respectively.
%The last equality follows from Eq.~(\ref{leretour}) and the fact that $\mathcal{A}(s)$
%is an operator on $\Ks$.
%As a consequence of Eq.~(\ref{eq:KAMtime}), Eqs.~(\ref{Vbhat}) yield %Eqs.~(\ref{eq:VStime})
%given in section~\ref{sub:time}

At the $n$-th iteration of the KAM algorithm, Eqs.~(\ref{TKH}) imply taking $\mathcal{K}\equiv \mathcal{K}\rnm$ and $\mathcal{V}\equiv \mathcal{V}_n$ in Eqs.~(\ref{Vbh}), hence $\mathcal{U}_\mathcal{\mathcal{H}}\equiv \mathcal{U}_{\mathcal{\mathcal{H}}\rnm}$ in Eq.~(\ref{Bst}).
%The result is Eqs.~(\ref{eq:VStime}) given in section~\ref{sub:time}.

\subsection{$\mathcal{U}_{\mathcal{H}\rn}(t,t_{0};s)$}
\label{sub:UH0n}
We show here that the operator $\mathcal{U}_{\mathcal{H}\rn}(t,t_{0};s)$ can be calculated according to Eq.~(\ref{eq:UH0na}), or equivalently Eq.~(\ref{eq:UH0nb}), in terms of $U_{H\rz}(t,t_{0})$ and the operators $\mathcal{S}_k(t,t_0;s)$ with $1 \leq k \leq n$ defined by Eq.~(\ref{eq:S}).
Indeed, applying Eq.~(\ref{eq:connectKH}) to the propagator of $\mathcal{K}\rn$ defined by Eqs.~(\ref{TKH}) yields
\BEA
\mathcal{U}_{\mathcal{H}\rn}(t,t_0;s)\!&=&\! {\sf T}_{t} \mathcal{U}_{\mathcal{K}\rn}(t,t_0) {\sf T}_{-t_0} \NN \\
&=&\! {\sf T}_{t} \mathcal{U}_{\mathcal{K}\rnm}(t,t_0)\mathcal{S}_n(t,t_0;s){\sf T}_{-t_0} \NN \\
&=&\! {\sf T}_{t} \mathcal{U}_{\mathcal{K}\rnm}(t,t_0){\sf T}_{-t_0}\mathcal{S}_n(t,t_0;s+t_0) \NN \\
&=&\! \mathcal{U}_{\mathcal{H}\rnm}(t,t_0;s)\mathcal{S}_n(t,t_0;s+t_0)  , \qquad
\label{UH0n}
\EEA
which proves Eq.~(\ref{eq:UH0na}).
Note that by construction $[\mathcal{K}\rnm, \overline{\mathcal{V}_{n}}^{\mathcal{K}\rnm}(s)]=0$, which is crucial for writing the second equality.
This commutation relation also allows to permute $\mathcal{U}_{\mathcal{K}\rnm}$ and $\mathcal{S}_n$ on the second line of Eqs.~(\ref{UH0n}), which then results in Eq.~(\ref{eq:UH0nb}).
 
\section{Averaging for pulse-driven systems}
\label{ap:pulse}
In this appendix, we consider the case of a time-dependent operator
$V(t)$  which, before some finite time $\ti$, is constant in time and commutes with the propagator of an Hamiltonian $H(t)$ on $\Hs$: 
\BES
\label{VUti}
\BEA
&&V(t)=V(\ti) \quad \forall t \leq \ti ,\label{Vti} \\
&&\left[V(\ti),U_{H}(t,t_0)\right]=0 \quad \forall t , t_0 \leq \ti . \label{Uti}
\EEA
\EES
After $\ti$ the dependence on time of $V(t)$ is arbitrary provided it is uniformly bounded.
We show that the operators $\overline{V}^{H}(t)$ and $\widehat{V}^{H}(t)$, defined by Eqs.~(\ref{eq:calVbhtime}), can be calculated as
\BES
\label{Vbhti}
\BEA
\overline{V}^{H}(t)\!\!\!&=&\!\!\!U_{H}(t,\ti) V(\ti) U_{H}^{\dagger}(t,\ti)  , \qquad \label{Vbti} \\
\widehat{V}^{H}(t)\!\!\!&=&\!\!\!\int_{\ti}^{t} \! \ud u \,U_{H}(t,u) \left[V(u)-\overline{V}^{H}(u)\right] U_{H}^{\dagger}(t,u) .\qquad \label{Vhti}
\EEA
\EES

We first prove Eq.~(\ref{Vbti}),  rewriting Eq.~(\ref{eq:calVbtime}) as
\BEA
\label{VbBst}
\overline{V}^{H}(t)\!&=&\!\lim_{T\rightarrow \infty} \frac{1}{T} 
\int_{t-T}^{t} \!\! \ud u  \, A(t,u)  , \
\EEA
where
$A(t,u)\equiv U_{H}(t,u) V(u) U_{H}^{\dagger}(t,u)$.
The propagator $U_{H}(t,u)$ can be decomposed into
$U_{H}(t,\ti)U_{H}(\ti,u)$.
If $u \leq \ti$ then $U_{H}(\ti,u)$ and $V(u)$ satisfy Eqs.~(\ref{VUti}) 
implying
\BEA
A(t,u)=U_{H}(t,\ti) V(\ti) U_{H}^{\dagger}(t,\ti) \quad \forall u \leq \ti  . \qquad \label{BstVbti}
\EEA
For $t\leq \ti$ the integration
domain in Eq.~(\ref{VbBst}) is such that $A(t,u)$ is given by Eq.~(\ref{BstVbti}), resulting thus in  Eq.~(\ref{Vbti}).
Notice that this latter reduces to
\BEA
\label{Vb0}
\overline{V}^{H}(t)=V(\ti) \quad \forall t\leq \ti .
\EEA
On the other hand, when $t > \ti$ there is a remaining integral over $u \in[\ti,t]$ which 
is bounded and independent of $T$, vanishing thus in the limit. 
The result given in Eq.~(\ref{Vbti}) follows.

We now turn to the proof of Eq.~(\ref{Vhti}) and write 
Eq.~(\ref{eq:calVhtime}) as
\BEA
\label{VhBst}
\widehat{V}^{H}(t)=\lim_{T\rightarrow \infty} \frac{1}{T} 
\int_{0}^{T} \!\!\! \ud t^{\prime}
\! \int_{t-t^{\prime}}^{t} \!\!\! \ud u  B(t,u)  , \
\EEA
where 
\BE
B(t,u)\equiv U_{H}(t,u)\left[V(u)-\overline{V}^{H}(u)\right]U_{H}^{\dagger}(t,u).
\EE  
The case $t \leq \ti$ is directly proven since $B(t,u)$, which is also the integrand in Eq.~(\ref{Vhti}), vanishes identically by Eq.~(\ref{Vb0}):
\BEA
\label{Vh0}
\widehat{V}^{H}(t)=0 \quad \forall t\leq \ti .
\EEA
For $t >\ti$ splitting the domains of integration in Eq.~(\ref{VhBst}) yields
\begin{eqnarray}
\!\!\!&&\!\!\!\widehat{V}^{H}(t)= \lim_{T\rightarrow \infty} \frac{1}{T}\left\{ \int_{0}^{t-\ti} \!\! \ud t^{\prime} \! \int_{t-t^{\prime}}^{\ti}\! \! \ud u B(t,u) \right. \NN\\
\!\!\!&&\!\!\! \left.
+  \int_{t-\ti}^{T} \!\! \ud t^{\prime} \! \int_{t-t^{\prime}}^{\ti} \!\! \ud u B(t,u) +\int_{0}^{T} \!\! \ud t^{\prime} \! \int_{\ti}^{t}\! \!\ud u B(t,u)\right\} \! .  \ \quad \label{Vhtibis}
\end{eqnarray}
The first double integral being bounded and independent of $T$ does not contribute in the limit whereas the second one vanishes  because $B(t,u)=0$ if $u \leq \ti$. The result of Eq.~(\ref{Vhti}) comes from the last double integral of Eq.~(\ref{Vhtibis}).

Finally, we show that if Eqs.~(\ref{VUti}) are satisfied with $V \equiv V_1$ and $H\equiv H\rz$, then Eqs.~(\ref{Vbhti}) hold with $V \equiv V_n$ and $H\equiv  H\rnm$ for any $n \geq 1$.
The case $n=1$ follows directly. 
We prove the case $n>1$ by induction, assuming Eqs.~(\ref{VUti}) are verified for $n-1$.
The operator $V_{n}(t)$ is obtained by Eq.~(\ref{eq:calVn+1}) as a sum of terms involving  $\widehat{V_{n-1}}^{H\rnmm}(t)$, which by Eq.~(\ref{Vh0}) 
is zero for $t \leq \ti$.
Hence $V_{n}(t)=0$ for $t \leq \ti$ so that Eqs.~(\ref{VUti}) hold for $n > 1$, which concludes the proof.
Notice that for $n>1$ Eq.~(\ref{Vbti}) yields $\overline{V_{n}}^{H\rnm}(s)=0$.

\bibliography{sudenKAM}
\end{document}